\documentstyle[12pt]{article}
\font\titlefont=cmbx10 scaled \magstep3
\begin{document}
\input{epsf}

\begin{flushright}
\vspace*{-2cm}
quant-ph/0003129  \\
CBPF-NF/015/00       \\  
Revised September 25, 2000  \\
\vspace*{2cm}
\end{flushright}

\begin{center}
{\titlefont Focusing Vacuum Fluctuations }
 
\vskip .3in
L.H. Ford\footnote{email: ford@cosmos2.phy.tufts.edu} \\
\vskip .1in
Institute of Cosmology,
Department of Physics and Astronomy\\
Tufts University\\
Medford, Massachusetts 02155\\
\vskip .2in
N.F. Svaiter\footnote{email:nfuxsvai@lafex.cbpf.br} \\ 
\vskip .1in
Centro Brasileiro de Pesquisas Fisicas-CBPF \\ 
Rua Dr. Xavier Sigaud 150\\ 
Rio de Janeiro, RJ 22290-180, Brazil \\
\end{center}

\vskip .2in
\begin{abstract}
The focusing of the vacuum modes of a quantized field by a parabolic mirror
is investigated. We use a geometric optics approximation to calculate the
energy density and mean squared field averages for scalar and electromagnetic
fields near the focus. We find that these quantities grow as an inverse power 
of the distance to the focus. There is an attractive Casimir-Polder force
on an atom which will draw it into the focus. Some estimates of the magnitude
of the effects of this focusing indicate that it may be observable.
\end{abstract}
\vskip .1in
 PACS categories: 03.70.+k, 34.20.Cf, 12.20.Ds, 04.62.+v.

\baselineskip=14pt

\section{Introduction}
\label{sec:intro}

The Casimir effect can be viewed as the reflection of vacuum fluctuations by 
mirrors. The presence of a reflecting boundary alters the modes of a
quantized field, and results in shifts in the vacuum expectation values
of quantities quadratic in the field, such as the energy density. Typically,
Casimir effects for massless fields may be estimated by dimensional analysis.
If $r$ is the distance to the nearest boundary, then the Casimir energy
density is typically of order $r^{-4}$ times a dimensionless constant. This
constant is usually of order $10^{-3}$ in four-dimensional spacetime. 
It is of course possible to find a much smaller result due to special
cancellations. For example, the Casimir energy density for a single, 
perfectly conducting plate is zero, even though the mean squared electric and
mean squared magnetic fields are separately nonzero.

 These typical results arise from calculations of specific geometries,
not from any general theorem. This leaves the possibility of exceptions,
where the energy density is much larger than would be expected on dimensional 
grounds. Indeed, one possible mechanism for amplification of vacuum
fluctuations has already been proposed \cite{Ford93,Ford98}. This mechanism
is based on the fact that the contribution of various parts of the frequency
spectrum to the Casimir effect is a highly oscillatory function 
\cite{Ford88,Hacyan}. The contributions of different ranges of frequency
almost, but not quite, completely cancel one another. The 
possibility that one can enhance the magnitude of the effect by altering
the reflectivity of the boundary in selected frequency ranges was discussed
in Refs.~\cite{Ford93,Ford98}. 

However, in this paper, we wish to propose a different mechanism for amplification
of vacuum fluctuations. This is the use of parabolic mirrors to create
anomalously large effects near the mirror's focus. It is well known in
classical physics that a parabolic mirror can focus incident rays which
are parallel to the mirror's axis. This means that a particular plane wave
mode becomes singular at the focus. One might wonder whether this classical
focusing effect of modes can produce large vacuum fluctuations near the focus. 
We will argue that the answer to this question is yes.

The outline of this paper is as follows: In Sect.~\ref{sec:Formalism}, the
essential formalism needed to compute mean squared field averages in the 
geometric optics approximation will be developed. It will be argued that
the dominant contributions will come from interference terms between different
reflected rays. In particular, expressions 
will be given for the case of two reflected rays from a single incident
ray in terms of the path length difference of the two reflected rays. In 
Sect.~\ref{sec:para} the specific case of parabolic mirrors will be studied,
and the condition for the existence of multiply reflected rays given.
It will be shown here that there is a minimum size required for a parabolic
mirror to produce large vacuum fluctuation focusing.
Section~\ref{sec:tech} deals with a couple of technical issues, including the
treatment of the apparently singular integrals which arise. In 
Sect.~\ref{sec:slightly}, we give explicit results in an approximation in
which the mirror is only slightly larger than the minimum size needed to
focus vacuum fluctuations. The possible experimental tests of these results
are discussed in Sect.~\ref{sec:obs}, and conclusions are given in
Sect.~\ref{sec:final}.

 Units in which $\hbar =c = 1$ will be used throughout this paper.
Electromagnetic quantites will be in Lorentz-Heaviside units. 

\section{Basic Formalism}
\label{sec:Formalism}

The approach which will be adopted in this paper is a geometric optics 
approximation. This approximation assumes that the dominant contribution
to the quantities which we calculate comes from modes whose wavelengths
are short compared to the goemetric length scales of the system. The
justification of the approximation will lie in a self-consistent calculation
leading to large contributions from short wavelength modes. At first sight,
it might seem that this approximation would always fail, and that only modes
whose wavelengths are of the order of the  goemetric length scales will
contribute significantly to quantities such Casimir energy densities.
However, there is a circumstance in which this intuition can fail. This is 
when there are two or more reflected rays produced by the same incident
beam. It then becomes possible to have an anomalously large interference 
term between these rays, as will be illustrated below. 

Let us first consider the case of a massless scalar field, $\varphi$.
Let the field operator be expanded in term of normal modes as
\begin{equation}
\varphi = \sum_{\bf k} (a_{\bf k}\, F_{\bf k} + 
          a^\dagger_{\bf k} \, F^*_{\bf k}) \, ,
\end{equation}
where $a^\dagger_{\bf k}$ and $a_{\bf k}$ are creation and annihilation
operators, and $F_{\bf k}$ are the mode functions. The formal vacuum expectation value
of $\varphi^2$ becomes 
\begin{equation}
\langle \varphi^2 \rangle_f =  \sum_{\bf k} \, |F_{\bf k}|^2 \,. \label{eq:phisq}
\end{equation}
In the absence of a boundary, the modes $F_{\bf k}$ are simply plane waves.
In the presence of the boundary, there are both incident and possibly one
or more reflected waves for each wave vector ${\bf k}$. Write the mode
function as
\begin{equation}
F_{\bf k} = f_{\bf k} + \sum_i  f^{(i)}_{\bf k} \, , \label{eq:mode}
\end{equation}
where $f_{\bf k}$ is the incident wave and the $f^{(i)}_{\bf k}$ are the
reflected waves. (Note that here ${\bf k}$ denotes the incident wavevector.)
 We may take all of these waves to be plane waves with
box normalization in a volume $V$, in which case
\begin{equation}
f_{\bf k} = \frac{1}{\sqrt{2 \omega V}} \, 
{\rm e}^{i({\bf k}\cdot{\bf x}-\omega t)} \,.             \label{eq:plane}
\end{equation}
The $f^{(i)}_{\bf k}$ take the same form, but with ${\bf k}$ replaced
by the appropriate wavevector for the reflected wave. 

If we now insert Eq.~(\ref{eq:mode}) into  Eq.~(\ref{eq:phisq}), we obtain
a sum involving both the absolute squares of the incident and the reflected
waves, and the various possible cross terms between the different waves:
\begin{equation}
\langle \varphi^2 \rangle_f = \sum_{\bf k} \left[|f_{\bf k}|^2 + 
 \sum_i |f^{(i)}_{\bf k}|^2 + \sum_i (f^*_{\bf k} f^{(i)}_{\bf k}
+f_{\bf k} {f^{(i)}}^*_{\bf k} ) 
+ \sum_{i \not= j} f^{(i)}_{\bf k} {f^{(j)}}^*_{\bf k}\right] \,.
\end{equation}
This quantity is divergent and needs to be renormalized by subtraction
of the corresponding quantity in the absence of boundaries. We will argue
in Sect.~\ref{sec:segments} that this is given by the above sum without 
the cross terms:
\begin{equation}
\langle \varphi^2 \rangle_0 =  \sum_{\bf k} (|f_{\bf k}|^2 + 
 \sum_i |f^{(i)}_{\bf k}|^2) \,.   \label{eq:phisqab}
\end{equation}
The renormalized expectation value is then given by the sum of cross terms
\begin{equation}
\langle \varphi^2 \rangle = 
\langle \varphi^2 \rangle_f -\langle \varphi^2 \rangle_0 = 
\sum_{\bf k} \left[\sum_i (f^*_{\bf k} f^{(i)}_{\bf k}
+f_{\bf k} {f^{(i)}}^*_{\bf k} ) 
+ \sum_{i \not= j} f^{(i)}_{\bf k} {f^{(j)}}^*_{\bf k} \right] \,.
\label{eq:phisqren}
\end{equation}
The renormalization which we employ is the usual one of defining the 
renormalized quantity to be the difference between the formal expectation
value with the mirror and that without it. Another way of expressing
the same prescription is to say that quantities such as 
$\langle \varphi^2 \rangle$ are only defined up to an additive constant,
and we choose the constant so that $\langle \varphi^2 \rangle \rightarrow 0$
at infinite distances from the mirror.

Let us examine a particular cross term:
\begin{equation}
T_{12} = \sum_{\bf k} (f^{(1)}_{\bf k} {f^{(2)}}^*_{\bf k} +
              f^{(2)}_{\bf k} {f^{(1)}}^*_{\bf k}) \,.
\end{equation}
Here $f^{(1)}_{\bf k}$ and $f^{(2)}_{\bf k}$ are both of the form of
Eq.~(\ref{eq:plane}), except with ${\bf k}$ replaced by ${\bf k}_1$ and
${\bf k}_2$, respectively. These might be two reflected waves, both 
corresponding to the same incident wavevector ${\bf k}$, but different
reflected wavevectors, ${\bf k}_1$ and ${\bf k}_2$. Only the direction,
not the frequency changes upon reflection, so
\begin{equation}
|{\bf k}_1| = |{\bf k}_2| = |{\bf k}| = \omega \, .
\end{equation}
We can now write
\begin{equation}
T_{12} = 2\, {\rm Re} \sum_{\bf k} \frac{1}{2 \omega V} \, 
{\rm e}^{i[{\bf k}_1 - {\bf k}_2)\cdot{\bf x}} \rightarrow
 \frac{1}{8 \pi^2} \int d^3 k 
  \frac{\cos[({\bf k}_1 - {\bf k}_2)\cdot{\bf x}]}{\omega} \,,
\end{equation}
where the infinite volume limit has been taken. The argument of the cosine
function is proportional to the difference in optical path lengths of the
two rays, $\Delta \ell$, so that
\begin{equation}
({\bf k}_1 - {\bf k}_2)\cdot{\bf x} = \omega \, \Delta \ell \, ,
\end{equation}
and hence
\begin{equation}
T_{12} = \frac{1}{8 \pi^2} \int d^3 k \, 
         \frac{\cos \omega \, \Delta \ell}{\omega} \, .
\end{equation}

Note that the integral in the above expression will diverge as 
$ (\Delta \ell)^{-2}$ in the limit that $\Delta \ell \rightarrow 0$.
Thus within the geometric optics approximation, we can obtain an
anomalously large contribution if there are two distinct reflected rays
with nearly the same optical path length. If this is the case, it provides
the self-consistent justification of the approximation. The dominant
contribution to the integral will come from modes with wavelength of the order 
of $\Delta \ell$; if this is small compared to all other length scales in the
problem, then the use of geometric optics should be a good approximation.

Normally one would expect Casimir effects to arise from modes whose
wavelengths are of the order of the length scales defined by the boundaries,
typically the distance to the nearest boundary. In this case, one would not
expect geometric optics to be a good approximation. Nonetheless, Schaden
and Spruch \cite{SS} have argued that one can often obtain reasonable
results from a semiclassical approximation involing a sum over periodic
classical orbits. Our use and justification of a geometric optics approximation
is perhaps more akin to that of Hawking \cite{Hawking} in his derivation
of black hole evaporation. There the modes which give the dominant
contribution to the Hawking radiation have very high frequencies when they
propagate through a collapsing star, and hence are accurately described by
geometric optics.

In this paper, we will examine the case of parabolic mirrors and show that
for points near the focus, there can be two reflected rays with nearly the
same path length. Their path lengths differ finitely from that of the 
incident ray. In this case, the dominant contribution to 
$\langle \varphi^2 \rangle$ comes from a single term of the form of $T_{12}$,
and we can write
\begin{equation}
\langle \varphi^2 \rangle \approx \frac{1}{8 \pi^2} \int d^3 k \, 
         \frac{\cos \omega \, \Delta \ell}{\omega} \, .  \label{eq:phisqren2}
\end{equation}
Note that the interference terms between the incident and the reflected rays
give a much smaller contribution because the $ \Delta \ell$ is much larger
for these terms.
We can also now write down expressions for several other quantities of
interest. These include $\langle {\dot \varphi}^2 \rangle$, where the dot 
denotes a time derivative, as well as the scalar field energy density
\begin{equation}
\rho_{scalar} = \frac{1}{2} \langle {\dot \varphi}^2  +
  |\nabla \varphi|^2  \rangle \approx \langle {\dot \varphi}^2 \rangle \,.
\end{equation}
In the last step we used the fact that 
\begin{equation}
| \dot f^{(i)}_{\bf k}| = | \nabla f^{(i)}_{\bf k}|
\end{equation}
for plane wave modes. We can also obtain renormalized expectation values for
electromagnetic field quantities, such as $\langle {\bf E}^2 \rangle$ and
$\langle {\bf B}^2 \rangle$, or the electromagnetic energy density
$\rho_{EM} = \frac{1}{2}(\langle {\bf E}^2 \rangle + \langle {\bf B}^2 \rangle)$.
Here ${\bf E}$ and ${\bf B}$ are the quantized electric and magnetic field
operators, respectively. The mode functions for these fields are of the form
of the right-hand-side of Eq.~(\ref{eq:plane}), except with an extra factor
of $\omega$ and a unit polarization vector. Thus, when we account for the
two polarizations of the electromagnetic field, we have
\begin{equation}
\langle {\bf E}^2 \rangle = \langle {\bf B}^2 \rangle =  \rho_{EM} 
= 2 \langle {\dot \varphi}^2 \rangle = 2 \rho_{scalar} 
= \frac{1}{4 \pi^2} \int d^3 k \, 
        \omega\, \cos \omega \, \Delta \ell \, .   \label{eq:Esqren}
\end{equation}

\section{Optics of Parabolic Mirrors}
\label{sec:para}

\subsection{Conditions for Multiply Reflected Rays}
\label{sec:cond}

\begin{figure}
\begin{center}
\leavevmode\epsfysize=8cm\epsffile{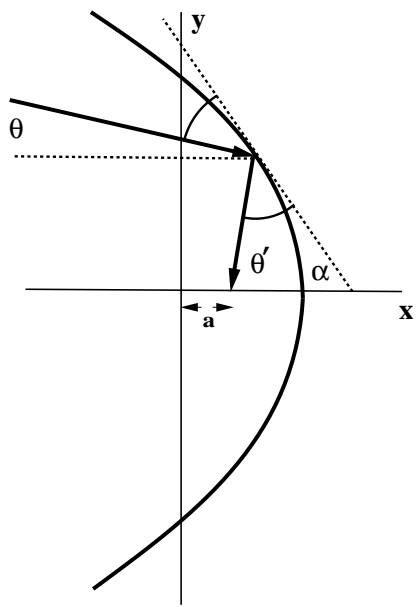}
\end{center}
\caption{A parabolic mirror has its focus at the origin. A ray incident at an angle 
$\theta$ with respect to the $x$-axis reflects off the mirror and arrives at a
point a distance $a$ from the origin at an angle $\theta'$. The line tangent
to the point of reflection is at an angle $\alpha$. }
\label{fig:mirror1}
\end{figure}

A parabolic mirror is illustrated in Fig.~1. The parabola described by 
\begin{equation}
x = \frac{b^2 -y^2}{2b}   \label{eq:para}
\end{equation}
has its focus at the origin, $x=y=0$. Consider a ray incident at angle
$\theta$ and reflected at angle $\theta'$ relative to the $x$-axis. Further
suppose that this ray reaches the $x$-axis at $x = a$, where $a \ll b$.
We wish to find the relationship between the angles $\theta$ and $\theta'$.
First note that
\begin{equation}
\theta = \theta' -\pi +2 \alpha \, ,  \label{eq:alpha1}
\end{equation}
where $\alpha$ is the angle of the tangent to the parabola at the point of
intersection. If we differentiate  Eq.~(\ref{eq:para}), we find
\begin{equation}
\frac{d y}{d x} = - \frac{b}{y_i} = - \tan \alpha \,,  \label{eq:alpha2}
\end{equation}
where here $y_i$ is the $y$-coordinate of the point of reflection. Note 
that the reflected ray is described by
\begin{equation}
y = \tan \theta' \,(x - a) \,.
\end{equation}
Combine this relation with Eq.~(\ref{eq:para}) to find
\begin{equation}
y_i = - \frac{b}{\tan \theta'} \left[ 1 \pm \sqrt{\sec^2 \theta'
      - 2\left(\frac{a}{b}\right) \tan^2 \theta'} \,\right] \,. \label{eq:yi}
\end{equation}
We expand this expression to first order in $a/b$ and note that for
$y_i > 0$, we need the minus sign before the square root. We then find
\begin{equation}
y_i \approx - \frac{b}{\tan \theta'} \left[ 1 -\sec \theta'
     + \left(\frac{a}{b}\right) \sin^2 \theta' \sec \theta' \right] \,.
\end{equation}
Now combine this result with  Eqs.~(\ref{eq:alpha1}) and (\ref{eq:alpha2})
to find, to first order in $a/b$, 
\begin{equation}
\theta = \frac{a\, \sin^3 \theta' \, \sec \theta'}{b (\sec \theta' -1)} \,.
                                                     \label{eq:theta}
\end{equation}
First, we note that $\theta \rightarrow 0$ as $a \rightarrow 0$ for fixed
$\theta'$. This is the expected result that all rays emanating from the focus
are reflected into parallel rays. Equation~(\ref{eq:theta}) is plotted
in Fig.~2. We see that for $a \not= 0$, there can be two reflected rays
for a given incident ray. However, one of the reflected rays always
corresponds to $\theta' > \pi/3$. Hence the mirror must subtend an angle
greater than $\pi/3$ as measured from the $x$-axis for this to happen.

\begin{figure}
\begin{center}
\leavevmode\epsfysize=6.5cm\epsffile{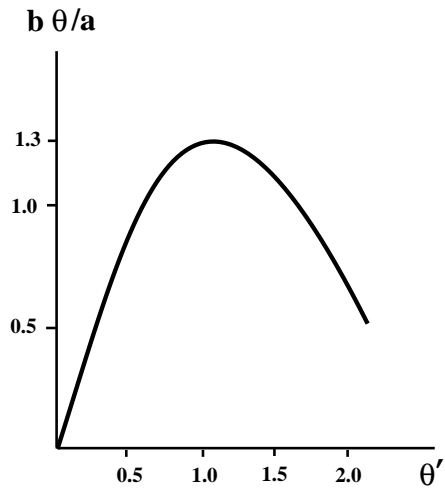}
\end{center}
\caption{The incident angle $\theta$ as a function of the angle $\theta'$ of the 
reflected ray. Both angles are measured in radians. Here it is assumed that 
the rays arrive near the focus
($a/b \ll 1$). For a single incident angle $\theta$, there can be two reflection
angles $\theta'$. The maximum of this curve occurs at $\theta' = \pi/3$. }
\label{fig:thetaplot}
\end{figure}

Our next task is to compute the difference in path lengths for these two
reflected rays. Consider first the distance $\ell$ which a particular ray 
travels
after it first crosses the line $x = a$. This distance can be broken into
two segments $s_1$ and $s_2$, as illustrated in Fig.~3. If $x_i$ is the 
$x$-coordinate of the reflection point, then
\begin{equation}
s_1 =\sqrt{(x_i -a)^2 + y^2_i}
\end{equation}
and 
\begin{equation}
s_2 = \frac{x_i -a}{\cos \theta} \,.
\end{equation}
We may now use Eqs.~(\ref{eq:para}) and (\ref{eq:yi}) to write
\begin{equation}
s_1 \approx \frac{ b}{\sin^2 \theta'} \,
\left[ 1-\cos \theta' -  \left(\frac{a}{b}\right) \sin^2 \theta' \right] \,.
\end{equation}
Similarly,
\begin{equation}
s_2 \approx x_i -a \approx \frac{ b\, \cos \theta'}{\sin^2 \theta'} \,
\left[ 1-\cos \theta' -  \left(\frac{a}{b}\right) \sin^2 \theta' \right] \,.
\end{equation}
Thus,
\begin{equation}
\ell = s_1 + s_2 = b -a (1 + \cos \theta') \,.
\end{equation}
If there are two distinct reflected rays with $\theta' =\theta_1'$ and
$\theta' =\theta_2'$, respectively, then the path length difference is
\begin{equation}
\Delta \ell = a(\cos \theta_1' - \cos \theta_2') \,. \label{eq:deltaL}
\end{equation}

\begin{figure}
\begin{center}
\leavevmode\epsfysize=8cm\epsffile{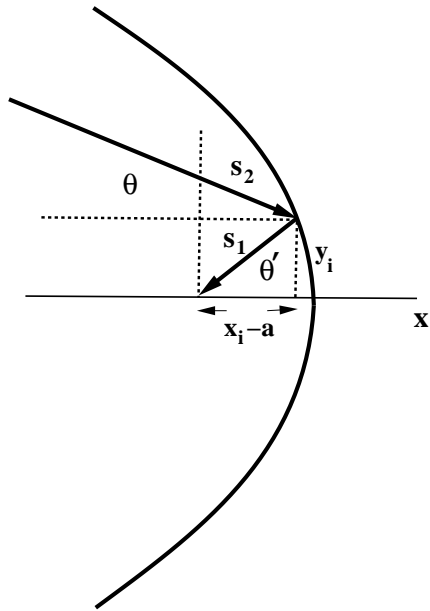}
\end{center}
\caption{The ray reflects from a point on the mirror with coordinates 
$(x_i,y_i)$ and then arrives at the point $(a,0)$. Here $s_2$ denotes the 
distance traveled by the ray from the $x=a$ line to the point of reflection,
and $s_1$ is the distance from that point back to $x=a$. }
\label{fig:Mirror2}
\end{figure}

\subsection{Parabola of Revolution}
\label{sec:rev}

In Section \ref{sec:para}, we dealt with the rays reflected from a parabola
in a plane. There are two ways to add on the third spatial dimension. One is to 
consider a parabolic cylinder and the other is to consider a parabola of 
revolution,
the surface formed by rotating a parabola about its symmetry axis. In the latter
case, one has an azimuthal angle $0 \leq \phi < 2\pi$. Thus 
Eq.~(\ref{eq:phisqren2}) becomes 
\begin{equation}
\langle \varphi^2 \rangle_{pr} =  \frac{1}{4 \pi^2} \int d\theta' \, 
\int_0^\infty d\omega    \, \omega    \cos \omega \, \Delta \ell =  
-\frac{1}{4 \pi^2 a^2} \int d\theta' \, 
       \frac{1}{(\cos \theta_1' -\cos \theta_2')^2} \,.\label{eq:phisqren3}
\end{equation}
Here we have evaluated 
\begin{equation}
\int_0^\infty d\omega \, \omega \cos \omega \, \Delta \ell = 
\lim_{\alpha \rightarrow 0} \int_0^\infty d\omega \, \omega 
      \cos \omega \Delta \ell \,\, {\rm e}^{-\alpha \omega} =
 -\frac{1}{\Delta \ell^2}  \, ,
\end{equation}
and then used Eq.~(\ref{eq:deltaL}).

We can do an analogous calculation for the various quantites given in
Eq.~(\ref{eq:Esqren}) to find, for example,
\begin{equation}
\langle {\bf E}^2 \rangle_{pr} = 
\frac{3}{2 \pi^2 a^4} \int d\theta' \, 
       \frac{1}{(\cos \theta_1' -\cos \theta_2')^4} \,.\label{eq:Esqren3}
\end{equation}
Here we have used 
\begin{equation}
\int_0^\infty d\omega \, \omega^3 \cos \omega \, \Delta \ell = 
\lim_{\alpha \rightarrow 0} \int_0^\infty d\omega \, \omega ^3
      \cos \omega \Delta \ell \, {\rm e}^{-\alpha \omega} =
 \frac{6}{(\Delta \ell)^4}  \, .
\end{equation}
Note that the integration on $\theta'$ in Eqs.~(\ref{eq:phisqren3}) and
(\ref{eq:Esqren3}) needs only to include positive values of $\theta'$.
If one reflects the ray illustrated in Fig.~1 through the $x$-axis in
the case of the parabola of revolution, one is going to a ray with the same
$\theta'$ but with $\phi \rightarrow \phi +\pi$.

\subsection{Parabolic Cylinder}
\label{sec:rcyl}

Another possible geometry in three space dimensions is that of the parabolic
cylinder. Let the cylinder be parallel to the $z$-direction. The wavevector
${\bf k}$ of the light rays now has a $z$-component, $k_z$, so that
\begin{equation}
\omega = \sqrt{\kappa^2 + k_z^2} \,,
\end{equation}
where $\kappa$ is the magnitude of the component of ${\bf k}$ in the $xy$ 
plane (perpendicular to the $z$-direction). If $s$ is a distance traveled
in the $xy$ plane, then the actual distance traveled is
 \begin{equation}
\sigma = s\, \frac{\omega}{\kappa} \,.
\end{equation}
Thus the difference in path lengths for a pair of reflected rays is 
$\omega \,\Delta \ell/\kappa$. We can modify Eq.~(\ref{eq:phisqren2}) 
to give an expression for the mean value of $\varphi^2$ near the focus of
a parabolic cylinder as
 \begin{equation}
\langle \varphi^2 \rangle_{pc} = \frac{1}{8 \pi^3} \int d^3 k \, 
         \frac{\cos (\omega^2 \, \Delta \ell/\kappa)}{\omega} \, .  
                      \label{eq:phisqpc}
\end{equation}
Similarly, the mean squared electric field is given by the analog of
Eq.~(\ref{eq:Esqren}):
\begin{equation}
\langle {\bf E}^2 \rangle_{pc} =  \frac{1}{4 \pi^3} \int d^3 k \, 
   \omega\, \cos (\omega^2 \, \Delta \ell/\kappa) \, .   \label{eq:Esqrenpc}
\end{equation}

The integrations in these two expressions are best done in cylindrical
coordinates, where $d^3 k = \kappa\, d\kappa \,d\theta'\, d k_z$. We can then
write
\begin{eqnarray}
\langle \varphi^2 \rangle_{pc} &=& \frac{1}{8 \pi^3} \int d\theta'
\int_0^\infty d\kappa \, \kappa\, {\rm Re} \left( {\rm e}^{i\kappa  \Delta \ell}
\int_{-\infty}^\infty \frac{d k_z}{\sqrt{\kappa^2 + k_z^2}} \,
{\rm e}^{i \Delta \ell \, k_z^2/\kappa} \right) \nonumber \\
&=& \frac{1}{8 \pi^3} \int d\theta' \,{\rm Re} \int_0^\infty d\kappa \, \kappa\, 
{\rm e}^{\frac{1}{2}i\kappa  \Delta \ell}\, 
   K_0\left(-\frac{1}{2}i\kappa  \Delta \ell\right) \,, \label{eq:phisqpc2}
\end{eqnarray}
where $K_0$ is a modified Bessel function, and in the last step we used
Formula 3.364.3 in Ref.~\cite{GR}. Next we use Formula 6.624.1 in the same
reference to write
\begin{eqnarray}
\lim_{\alpha \rightarrow \beta} \int_0^\infty dx \, x\, {\rm e}^{-\alpha x}
K_0(\beta x) &=& \lim_{\alpha \rightarrow \beta} \; 
\left(\frac{1}{\alpha^2-\beta^2}
\, \left\{ \frac{\alpha}{\sqrt{\alpha^2-\beta^2}} 
\ln \left[\frac{\alpha}{\beta} + \sqrt{ \left(\frac{\alpha}{\beta}\right)^2
 -1} \right] -1 \right\} \right) \nonumber \\
 &=& \frac{1}{3 \beta^2} \,. \label{eq:ident}
\end{eqnarray}
We can combine this last result with Eqs.~(\ref{eq:phisqren3}) and 
(\ref{eq:phisqpc2}) to find an expression for $\langle \varphi^2 \rangle_{pc}$.
However, we need to account for the fact that here, unlike the parabola
of revolution, the integration on $\theta'$ runs over negative values.
This can be done by introducing a factor of two (corresponding to the 
contributions of the upper and lower halfs of the cylinder) and writing
\begin{equation}
\langle \varphi^2 \rangle_{pc} = 
                \frac{4}{3 \pi} \, \langle \varphi^2 \rangle_{pr} \,.
\end{equation}
Similarly, we can write
\begin{eqnarray}
\langle {\bf E}^2 \rangle_{pc} &=&  \frac{1}{4 \pi^3} \int d\theta'
\int_0^\infty d\kappa \, \kappa^2 \, \frac{d}{d\, \Delta \ell} \,
\int_{-\infty}^\infty \frac{d k_z}{\sqrt{\kappa^2 + k_z^2}} 
\sin \left(\frac{\kappa^2 + k_z^2}{\kappa}\, \Delta \ell \right) \nonumber \\
&=& \frac{1}{4 \pi^3} \int d\theta' \,  \frac{d}{d\, \Delta \ell} \, {\rm Im} 
\int_0^\infty d\kappa \, \kappa^2 \, {\rm e}^{\frac{1}{2}i\kappa  \Delta \ell}\, 
           K_0\left(-\frac{1}{2}i\kappa  \Delta \ell\right) \,.
\end{eqnarray}
If we differentiate Eq.~(\ref{eq:ident}) with respect to $\alpha$ before taking
the limit, we may show that
\begin{equation}
\lim_{\alpha \rightarrow \beta} \int_0^\infty dx \,x^2 \, {\rm e}^{-\alpha x}
K_0(\beta x) =\frac{4}{15 \beta^3} \,. 
\end{equation}
This last identity and Eq.~(\ref{eq:Esqren}) may be used to show that
\begin{equation}
\langle {\bf E}^2 \rangle_{pc} = 
                \frac{16}{15 \pi} \, \langle {\bf E}^2 \rangle_{pr} \,.
\end{equation}
Thus the results for the parabolic cylinder are related to those for the
parabola of revolution by a numerical factor somewhat less than unity.

\section{Further Technical Issues}
\label{sec:tech}

\subsection{Reflected Rays from Line Segments}
\label{sec:segments}

In this subsection, we examine the problem of the reflection of rays from 
a pair of attached line segments, as illustrated in Fig.~4. The purpose
of this exercise is twofold: First, it will lead to the justification of
the renormalization prescription used in writing down Eq.~(\ref{eq:phisqab}).
Second, it will reveal that integrals such as those in Eqs.~(\ref{eq:phisqren})
and (\ref{eq:Esqren}) should involve an integration over $\theta'$, the angle
of the reflected wave, rather than $\theta$, the angle of the incident wave.

\begin{figure}
\begin{center}
\leavevmode\epsfysize=6.5cm\epsffile{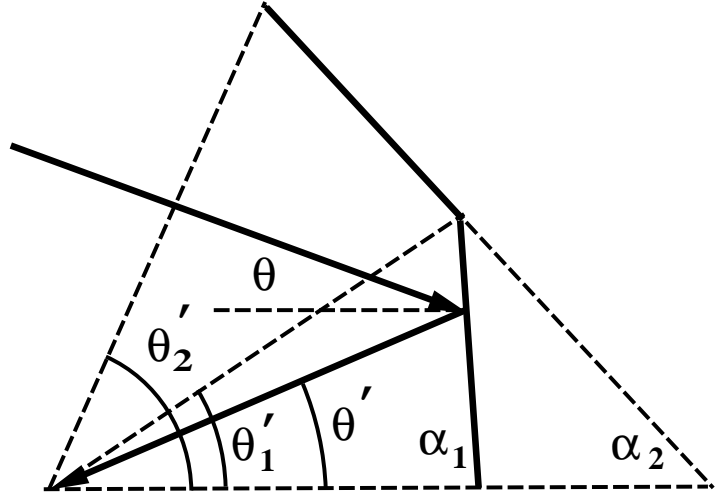}
\end{center}
\caption{Two flat mirror segments are aligned at angles $\alpha_1$ 
and $\alpha_2$,
respectively, and subtend angles $\theta_1'$ and $\theta_2' - \theta_1'$
from the point of interest. An incident ray has angle $\theta$ with respect 
to the $x$-axis, whereas the reflected ray has angle $\theta'$. }
\label{fig:flatsegs}
\end{figure}

First consider the case $0 < \theta' < \theta_1'$, so the ray reflects
from the lower segment oriented at angle $\alpha_1$ relative to the horizontal.
Here $\theta = \theta' + 2 \alpha_1 -\pi $ and hence
\begin{equation}
 2 \alpha_1 -\pi < \theta <  \theta_1' + 2 \alpha_1 -\pi \,.
\end{equation}
Now consider the case where the ray reflects from the upper segment, so
 $\theta_1' < \theta' < \theta_2'$ and $\theta = \theta' +2 \alpha_2 -\pi $. 
Here
\begin{equation}
\theta_1' + 2 \alpha_2 -\pi < \theta <  \theta_2' + 2 \alpha_2 -\pi \,.
\end{equation}

Note that the range of $\theta'$ is $\Delta \theta' = \theta_2'$ whereas
the range of $\theta$ is 
\begin{equation}
\Delta \theta = \theta_2' + 2 (\alpha_2 - \alpha_1) < \theta_2' \,.
\end{equation}
However, for $\theta$ in the range $\theta_1' + 2 \alpha_2 -\pi < \theta <  
\theta_1' + 2 \alpha_1 -\pi$, there are two reflected rays for each incident
ray. This is a range of $\delta \theta = 2 (\alpha_2 - \alpha_1)$, and we
have
\begin{equation}
 \delta \theta + \Delta \theta = \Delta \theta' \,.
\end{equation}
 
Although $\theta'$ runs over a larger range than does $\theta$, we can think 
of this larger range as counting the multiple reflected rays that can result
from an incident ray with a given value of $\theta$. This conclusion will
continue to hold if we have more than two straight line segments. We can 
approximate any curve by a sequence of line segments. In general, the angle
$\Delta \theta'$ subtended by the curve differs from the range of angle of
incident rays, $\Delta \theta$, and if the curve is convex toward the point
of interest, $\Delta \theta < \Delta \theta'$. At first sight, one might think 
that an emmeration of the independent modes should involve an integration
over $\theta$. This, however, fails to account for the multiple reflected rays,
which are correctly counted if we instead integrate on $\theta'$. 

As we vary $\theta$ through its range of $ -\pi < \theta \leq \pi$ (Note
that here $\theta$ increases in the clockwise direction.), we have six
possibilities:
\begin{eqnarray*}
&- \pi < \theta <  2 \alpha_1 -\pi& 
 \qquad {\rm incident\; ray\; only} \\
 &2 \alpha_1 -\pi < \theta <  \theta_1' + 2 \alpha_2 -\pi& \qquad
 {\rm 1 \; reflected \; ray} \\
&\theta_1' + 2 \alpha_2 -\pi < \theta < \theta_1' + 2 \alpha_1 -\pi& \qquad
 {\rm 2 \; reflected \; rays} \\
&\theta_1' + 2 \alpha_1 -\pi < \theta < \theta_2' + 2 \alpha_2 -\pi& \qquad
 {\rm 1 \; reflected \; ray} \\
&\theta_2' + 2 \alpha_2 -\pi < \theta < \pi - \theta_2'& \qquad
{\rm incident\; ray\; only} \\
&\pi - \theta_2' < \theta \leq \pi& \qquad
{\rm no \; rays.}
\end{eqnarray*}
In the latter case the incident ray fails to reach the point of interest because
it is blocked by the mirror. Note, however, that the reflected rays exactly
compensate for the missing incident rays in the sense that if we add up a 
weighted sum of the angle ranges with reflected rays, it is equal to the range
with no rays. This observation is the justification for Eq.~(\ref{eq:phisqab}).
The number of incident plus reflected rays in the presence of the boundary is
the same as the number of incident rays in its absence. 

One might ask whether it is important also to include the interference terms
between the multiply refelcted rays. After all, the dominant contribution near 
the focus of a parabolic mirror comes from such an interference term. It
is indeed true that if one wishes to compute a quantity such as  
$\langle {\bf E}^2 \rangle$ in the geometry of Fig.~4, we would need to include
the interference terms. However, one should not expect to obtain an anomalously
large result, but rather one of order $r^{-4}$, where $r$ is the distance to
the nearest boundary. This follows from the fact that the formula for
$\Delta \ell$, the analog of Eq.~(\ref{eq:deltaL}) will be of the form of
a product of $r$ times a dimensionless angular dependent function.

\subsection{Evaluation of Singular Integrals}
\label{sec:sing}

We have derived expressions, such as Eqs.~(\ref{eq:phisqren3}) and 
(\ref{eq:Esqren3}), for renormalized quantities near the focus of a parabolic 
mirror. Recall that we are dealing with a situation where there are two 
reflected rays for a single incident ray. Here $\theta_1'$ is the angle of 
one of these rays, and the angle of the other,  $\theta_2'$ is understood
to be a function of $\theta_1'$. However, the integrals in question are
singular at the point that $\theta_1' = \theta_2'$. The singularity may
be removed by an integration by parts \cite{Davies,FJL,WF99}. We rewrite the 
integrand using relations such as
\begin{equation}
\frac{1}{x^2} = - \frac{1}{2} \frac{d^2}{d x^2} \ln x^2
\end{equation}
and 
\begin{equation}
\frac{1}{x^4}= - \frac{1}{12}\, \frac{d^4}{d x^4} \ln x^2 \, .
\end{equation}
Next we perform repeated integrations by parts until we have only an integral
with a logarithmic singularity in the integrand, plus possible surface terms.
Thus, for example,
\begin{equation}
\int dx\, \frac{f(x)}{x^2} = - \frac{1}{2}\int dx\,\ln x^2\,
\frac{d^2 f(x)}{d x^2} \, ,
\end{equation}
and 
\begin{equation}
\int dx\, \frac{f(x)}{x^4} = -\frac{1}{12}\int dx\,\ln x^2\,
\frac{d^4 f(x)}{d x^4} \, ,
\end{equation}
provided that the function $f(x)$ is regular at $x=0$ and the surface terms 
vanish. This procedure is a generalization of the notion of a principal value
to cases of higher order poles. 

In our case, the integration on $\theta_1'$ ranges over those values of
${\theta}'$ for which there are multiple reflected rays. Within the geometric
optics approximation, the integrand would seem to drop precipitously to zero
at the end point of this interval. If one were to go beyond this approximation,
the sudden drop would be smeared out over an interval corrresponding to about
one wavelength. Thus we can think of our integrand as being an approximation
to a function which, along with its derivatives, vanishes smoothly at the
endpoints. If so, then we can ignore the surface terms. In any
case, we will here make the assumption that the surface terms can be ignored.
The integrand falling smoothly to zero can arise from more than one physical
cause. One is diffraction effects at the edge of the mirror, as noted above.
Another way to enforce this behavior is to consider a mirror in which the
reflectivity falls smoothly from near unity to zero as the edge of the mirror
is approached.

\section{Results for Mirrors Slightly Larger than the Critical Size}
\label{sec:slightly}

As we found above (See Fig.~2.), there is a critical size which a parabolic 
mirror must have before we find large vacuum effects near the focus. The
critical case is that of a mirror which subtends an angle of $\pi/3$ in
either direction from the axis of symmetry (the $x$-axis in Fig.~1). In
order to evaluate the integrals in Eqs.~(\ref{eq:phisqren3}) and
(\ref{eq:Esqren3}), we need to solve Eq.~(\ref{eq:theta}) for $\theta'$ in
terms of $\theta$, and then express one root $\theta_2'$ as a function
of the other, $\theta_1'$. In general, this is difficult to do in closed
form. There is, however, one case in which an analytic approximation is
possible. This is when the size of the mirror is only slightly greater than
the critical value. Let the angle subtended by the mirror be $\pi/3 + \xi_0$,
where $\xi_0 \ll 1$. In this case, we can expand the needed quantities in terms
of power series. Note that now both roots for $\theta'$ will be close to
$\pi/3$, so let $\theta' = \pi/3 +\xi$ and expand Eq.~(\ref{eq:theta}) in
powers of $\xi$ to find (This and other calculations in this section were 
performed using the computer algebra program MACSYMA.)
\begin{equation}
\theta = \frac{3\sqrt{3}}{4} -\frac{3\sqrt{3}}{4} \xi^2 +\frac{1}{4}\xi^3
+ \frac{3\sqrt{3}}{16} \xi^4 -\frac{1}{16}\xi^5 -\frac{11\sqrt{3}}{480} \xi^6
+ \cdots \,.  \label{eq:theta2}
\end{equation}
Let $\theta_1' = \pi/3 + \xi_1$ and $\theta_2' = \pi/3 + \xi_2$. Assume
a power series expansion for $\xi_2$ in terms of $\xi_1$. Next we equate
the right-hand-side of Eq.~(\ref{eq:theta2}) with $\xi=\xi_1$ to that with
$\xi=\xi_2$ and iteratively solve for the coefficients in the expansion of
$\xi_2$. The result is
\begin{equation}
 \xi_2 = - \xi_1 + \frac{\sqrt{3}}{3} \xi_1^2 -\frac{1}{27}\xi_1^3
+ \frac{35\sqrt{3}}{972}\xi_1^4 -\frac{97}{2916}\xi_1^5 + \cdots \,. 
                                               \label{eq:xi2}
\end{equation}

Our next task is to use this expansion to compute the integrands in
Eqs.~(\ref{eq:phisqren3}) and (\ref{eq:Esqren3}). First rewrite these
expressions as
\begin{equation}
\langle \varphi^2 \rangle_{pr} =  
-\frac{1}{8 \pi^2 a^2} \int d \,\xi_1
\frac{1}{[\cos (\frac{\pi}{3}+\xi_1) -\cos(\frac{\pi}{3}+\xi_2) ]^2} \,,
 \end{equation}
and 
\begin{equation}
\langle {\bf E}^2 \rangle_{pr} = 
\frac{3}{4 \pi^2 a^4} \int d \,\xi_1
\frac{1}{[\cos (\frac{\pi}{3}+\xi_1) -\cos(\frac{\pi}{3}+\xi_2) ]^4} \,.
\end{equation}
Again we must emphasize that the integrands in these expressions are the
approximate forms away from the end points of the integrations, but should
actually vanish at the end points. Next we replace these expressions by the 
forms obtained by the integratios by parts described above, where the surface 
terms are assumed to vanish. After the integrations by parts, we can
recognize that the dominant contributions to the integrals come from the 
interval $[-\xi_0, \xi_0]$ and write
\begin{equation}
\langle \varphi^2 \rangle_{pr} =  
 \frac{1}{16 \pi^2 a^2} \int_{-\xi_0}^{\xi_0} d \,\xi_1
 \ln{\xi_1}^2 \,    \frac{d^2}{d \xi_1^2} \,
  \left\{ \frac{\xi_1^2 }
  {[\cos (\frac{\pi}{3}+\xi_1) -\cos(\frac{\pi}{3}+\xi_2) ]^2}\right\}  \,,
\label{eq:phisqren5}
\end{equation}
and 
\begin{equation}
\langle {\bf E}^2 \rangle_{pr} = 
 -\frac{3}{48 \pi^2 a^4} \int_{-\xi_0}^{\xi_0} d \,\xi_1
 \ln{\xi_1}^2 \,    \frac{d^4}{d \xi_1^4} \,
 \left\{ \frac{\xi_1^4 }
       {[\cos (\frac{\pi}{3}+\xi_1) -\cos(\frac{\pi}{3}+\xi_2) ]^4}\right\}  \,.
\label{eq:Esqren5}
\end{equation}
Note that we have introduced a factor of $\frac{1}{2}$ to compensate for
overcounting of pairs of reflected rays. 

Next we use Eq.~(\ref{eq:xi2}) to write
\begin{equation}
\frac{\xi_1^2 }
     {[\cos (\frac{\pi}{3}+\xi_1) -\cos(\frac{\pi}{3}+\xi_2) ]^2} 
     = A_0 + A_1 \xi_1 + A_2 \xi_1^2 + \cdots  \,,
\end{equation}
and 
\begin{equation}
\frac{\xi_1 ^4 }
     {[\cos (\frac{\pi}{3}+\xi_1) -\cos(\frac{\pi}{3}+\xi_2) ]^4} 
     = B_0 + B_1 \xi_1 + B_2 \xi_1^2 + B_3 \xi_1^3 + B_4 \xi_1^4 +\cdots \,.
\end{equation} 

We see, that to leading order in $\xi_0$, the dominant contribution to
$\langle \varphi^2 \rangle$ comes from the coefficient $A_2$, which is
given by
\begin{equation}
A_2 = \frac{23}{324} \,.
\end{equation}
This leads to our final result
\begin{equation}
\langle \varphi^2 \rangle_{pr} \approx - \frac{23}{648 \pi^2 a^2}\, \xi_0
(1 - \ln\xi_0) + O(\xi_0^2 \ln\xi_0) \,.
\end{equation}
Similarly, the leading contribution to $\langle {\bf E}^2 \rangle_{pr}$ comes
from
\begin{equation}
B_4 = \frac{4051}{2^4 3^8 5} \,,
\end{equation}
and is
\begin{equation}
\langle {\bf E}^2 \rangle_{pr} \approx \frac{4051}{2^2 3^7 5 \pi^2 a^4}\, \xi_0
(1 - \ln\xi_0) + O(\xi_0^2 \ln\xi_0)  \approx \frac{9.38 \times 10^{-3}}{ a^4}
 \, \xi_0 (1 - \ln\xi_0) \,.\label{eq:E2pr}
\end{equation}

First we note that the leading contributions to both quantites diverge as
$a \rightarrow 0$, that is, as one approaches the focus. This provides the
justification of the geometric optics approximation. The modes which give the 
dominant contribution are those whose wavelengths are of order $a$, small
enough that geometric optics is valid. Next we note that 
$\langle \varphi^2 \rangle_{pr}$ diverges negatively, but 
$\langle {\bf E}^2 \rangle_{pr}$ and the energy density for the scalar and 
electromagnetic fields diverge positively.

The above results apply in the case of a parabola of revolution; for the
case of a parabolic cylinder we have
\begin{equation}
\langle \varphi^2 \rangle_{pc} \approx - \frac{23}{486 \pi^3 a^2}\, \xi_0
(1 - \ln\xi_0) \, ,
\end{equation}
and
\begin{equation}
\langle {\bf E}^2 \rangle_{pc} \approx \frac{16204}{ 3^8 5^2 \pi^3 a^4}\, \xi_0
(1 - \ln\xi_0)  \approx \frac{3.18 \times 10^{-3}}{ a^4}  
         \, \xi_0 (1 - \ln\xi_0)  \,.\label{eq:E2pc}
\end{equation}

Note that all of the results in this section depend upon what is happening 
in a thin band centered on $\theta' = \pi/3$. The remainder of the mirror,
that for which $\theta' < \pi/3 -\xi_)$, does not even have to be present.

\section{Observable Consequences?}
\label{sec:obs}

Now we face the question of whether the amplified vacuum fluctuations
are actually observable. The calculations given above indicate that the
energy density and squared fields are singular at the focus of a perfectly
reflecting parabolic mirror. However, the approximation of perfect reflectivity
must break down at frequencies higher than the plasma frequency of the material
in question . So long as the plasma wavelength $\lambda_P$ is short compared 
to the size of the mirror, there is an intermediate regime in which geometric 
optics is valid. We simply must restrict the use of the  geometric optics
results  to values of $a$ larger than $\lambda_P$. 

The quantity which is most easily observable is $\langle {\bf E}^2 \rangle$,
as it is linked to the Casimir force on an atom or a macroscopic particle.
If the atom or particle has a static polarizability $\alpha$, then the
interaction energy with a boundary is
\begin{equation}
V = - \frac{1}{2}\, \alpha\, \langle {\bf E}^2 \rangle \,. \label{eq:CP}
\end{equation}
Here we are assuming that the modes which give the dominant contribution
to $\langle {\bf E}^2 \rangle$ have frequencies below that at which a dynamic 
polarizability must be used. For a perfectly conducting parallel plate,
\begin{equation}
\langle {\bf E}^2 \rangle_{\rm plate} = \frac{3}{16\, \pi^2\, z^4}
 \approx \frac{1.90 \times 10^{-2}}{ z^4} \, , \label{eq:E2plate}
\end{equation}
where $z$ is the distance to the plate. If we insert this expression into
 Eq.~(\ref{eq:CP}), then the result is the Casimir-Polder potential \cite{CP}
for the interaction of an atom in its ground state with the plate. It is a 
good approximation when $z$ is large compared to the wavelength associated 
with the transition between the ground state and the first excited state.
The $1/z^4$ distance dependence of the Casimir-Polder potential was 
experimentally confirmed by Sukenik {\it et al} \cite{Sukenik}. If we compare
Eq.~(\ref{eq:E2plate}) with Eq.~(\ref{eq:E2pr}) or Eq.~(\ref{eq:E2pc}), we
see that the mean squared electric field near the focus of a parabolic
mirror is only slightly less than that at the same distance from a flat
plate. Given that the latter has actually been observed, it is possible that
the inhanced fluctuations near the focus are also observable by techniques
similar to those by Sukenik {\it et al}.

The basic method used in the Sukenik {\it et al} experiment is to look for the 
effects of the deflection of a beam of atoms as it passes near a pair of
plates. We can give a general estimate of the size of this type of deflection
which applies whenever there is a mean squared electric field which varies
as the inverse fourth power of a length scale. Let
\begin{equation}
\langle {\bf E}^2 \rangle = \frac{\Lambda}{a^4} \, ,
\end{equation}
where $a$ is the length scale and $\Lambda$ is a dimensionless constant. We
assume that an atom has an interaction of the form of  Eq.~(\ref{eq:CP}).
The resulting force, $F = - \partial V/\partial a$, will cause a deflection
$\Delta a$ in the atom's position in a time $t$, where
\begin{equation}
\frac{\Delta a}{a} = 0.25\, \left(\frac{\Lambda}{10^{-3}} \right)\, 
\left(\frac{\alpha}{\alpha_{Na}} \right)\,
 \left(\frac{m_{Na}}{m} \right)\, \left(\frac{1 \mu {\rm m}}{a} \right)^6\, 
\left(\frac{t}{10^{-3} s} \right)^2 \,.
\end{equation}
Here $m_{Na}=3.8 \times 10^{-23} {\rm gm}$ and 
$\alpha_{Na}=3.0\times 10^{-22} {\rm cm}^3$ denote the mass and polarizability 
of the sodium atom, respectively. (Note that polarizability in the 
Lorentz-Heaviside which we use is $4 \pi$ times that in Gaussian units.) 
If $t$ is of order $10^{-3} s$ (the time needed
for an atom with a kinetic energy of order 300K to travel a few centimeters),
and $z$ is of order $1 \mu {\rm m}$, the fractional deflection is significant.
Recall that in our case
\begin{equation}
\Lambda = \left\{ \begin{array}{ll}
  9.38 \times 10^{-3}\, \xi_0 (1 - \ln\xi_0)\,, & 
                                             \mbox{parabola of revolution}\\
  3.18 \times 10^{-3}\, \xi_0 (1 - \ln\xi_0)\,, & \mbox{parabolic cylinder}.
  \end{array}    \right. 
\end{equation}             
Thus it may be possible to observe the force on atoms near the focus.

Another possible way to observe this force might be to levitate the atoms
in the Earth's gravitational field. (A rather different form of levitation
by Casimir forces was proposed in Ref.~\cite{Ford98}.) If one equates the
force on atom at a distance $a$ from the focus to its weight, the result
can be expressed as
\begin{equation}
a = \left( \frac{2 \Lambda \alpha}{m g} \right)^\frac{1}{5}
= 0.55 \mu {\rm m} \, \left[ \left(\frac{\Lambda}{10^{-3}} \right)\, 
\left(\frac{\alpha}{\alpha_{Na}} \right)\,
 \left(\frac{m_{Na}}{m} \right) \right]^\frac{1}{5} \,.
\end{equation} 
Given that this formula applies for $a > \lambda_P$ and that 
$\lambda_P \approx 0.1 \mu {\rm m}$ for many metals, it seems possible that
levitation near the focus is possible. Of course, atoms will only be
trapped if their temperature is sufficiently low. The required temperature 
can be estimated by setting the thermal energy $\frac{3}{2} k T$ equal to the 
magnitude of the potential energy $V$. The result is
\begin{equation}
T = 2 \times 10^{-5} K\, \left(\frac{\Lambda}{10^{-3}} \right)\, 
\left(\frac{\alpha}{\alpha_{Na}} \right)\,
\left(\frac{0.1 \mu {\rm m}}{a} \right)^4 \,.
\end{equation}
Thus for $a$ of the order of a few times $0.1 \mu {\rm m}$, the required
temperature is larger than the temperatures of the order of $10^{-7} K$
which have already been achieved for laser cooled atoms \cite{Phillips,WPW}.

Another possibility might be the use of atom interferometry.
Atoms traveling for a time $t$ parallel to and
near the focus of a parabolic cylinder will acquire a phase shift of
\begin{equation}
\Delta \phi = \frac{t}{2}\, \alpha\, \langle {\bf E}^2 \rangle_{pc} 
= 0.14 \,\left(\frac{\alpha}{\alpha_{Na}} \right)\,
\left(\frac{1 \mu {\rm m}}{a} \right)^4 \,\left(\frac{t}{10^{-3} s} \right) \,
  \xi_0 (1 - \ln\xi_0) \,.
\end{equation}
If it is possible to localize the atoms to within a few $\mu {\rm m}$ of the
focus, then the accumulated phase shift for reasonable flight times would
seem to be within the currently attainable sensitivities of the order of 
$10^{-4}$ radians \cite{WPW}.

\section{Discussion and Conclusions}
\label{sec:final}

In this paper we have argued that a parabolic mirror is capable of focusing
the vacuum modes of the quantized electromagnetic field and creating
large physical effects near the mirror's focus. Just as the mirror can
focus a beam of light, it can focus something even in the absence of incoming
light. This might be dubbed ``focusing a beam of dark'' \cite{PD}.
The manifestation of this focusing is a growth in the energy density
and mean squared electric field as the focus is approached. In the idealized
case of a perfectly reflecting mirror, these quantities diverge as the inverse
fourth power of the distance from the focus. For a real mirror, the growth is
expected to saturate at distances of the order of the plasma wavelength of
the mirror. 

The most readily observable consequence of the focused vacuum fluctuations
is enhanced Casimir forces on atoms or other particles near the focus. The
sign of the force is such as to draw particles into the vicinity of the
focus. Estimates given in the previous section indicate that the magnitude
of this effect may be large enough to be observable.

The calculations presented in this paper were based on the geometric optics
approximation in which only short wavelenth modes are considered. The 
justification of this approximation is self-consistency: the large effects
near the focus can only come from the short wavelength modes for which the
approximation should be a good one. Nonetheless, in future work it will
be of interest to go beyond the geometric optics approximation. This should
allow one to check the validity of the assumption made in Sect.~\ref{sec:sing}
that the surface terms can be ignored.

In order to simplify the calculations, we made two restrictions on the
geometry. The first is that we have assumed that the point at which
the mean squared field quantites are measured lies on the symmetry axis
of the parabola (the $x$-axis). The second is that the mirror be only slightly
larger than the critical angle of $\pi/3$ at which vacuum focusing begins.
(This is the assumption that $\xi_0 \ll 1$, made in Sect.~\ref{sec:slightly}.)
It is of interest to remove both of these restriction, which we hope to do
in a future work.

\vspace{0.5cm}

{\bf Acknowledgement:}  We would like to thank Paul Davies for helpful
discussions. This work was supported in part by the National
Science Foundation under Grant PHY-9800965.

\end{document}